\documentclass[aps,prd,10pt, twocolumn, superscriptaddress, nofootinbib]{revtex4}
\usepackage{mathrsfs, amssymb, amsmath}  
\usepackage{epsfig, cancel}
\usepackage{latexsym}
\usepackage{natbib, comment}
\usepackage{url}
\usepackage{dcolumn}
\usepackage{multirow}
\usepackage{color}
\usepackage{cancel}
\usepackage{soul}
\usepackage[normalem]{ulem}
\usepackage{amsfonts,amssymb,amsmath, txfonts}
\usepackage{graphicx,epsfig}
\usepackage{psfrag}
\usepackage{hyperref}
\hypersetup{colorlinks=true}
\usepackage{mathtools}
\usepackage{enumitem}
\usepackage{float}
\usepackage[dvipsnames]{xcolor}
\usepackage{xcolor}
\hypersetup{ linktoc=all,
    colorlinks, linkcolor={brightpink},
    citecolor={blue}, urlcolor={blue}
}
\definecolor{rosy}{RGB}{230,235,252}
\definecolor{myframetitle}{RGB}{90,89,170}
\definecolor{myblocktitle}{RGB}{140,185,249}
\definecolor{mytitle}{RGB}{10,80,26}

\definecolor{darkgreen}{RGB}{27,130,45}
\definecolor{darkblue}{rgb}{0,0,0.3}
\definecolor{darkred}{rgb}{0.7,0,0}

\definecolor{light gray}{RGB}{220,220,220}
\definecolor{dark purple}{RGB}{108,0,217}
\definecolor{pink}{RGB}{190,20,100}
\definecolor{orang}{RGB}{193,63,0}
\definecolor{green}{RGB}{11,98,17}
\definecolor{darkpink}{RGB}{153,0,76}
\definecolor{bluegreen}{RGB}{0,102,102}
\definecolor{greenlagan}{RGB}{0,102,0}
\definecolor{redgreen}{RGB}{102,102,0}
\definecolor{Redgreen}{RGB}{153,76,0}
\definecolor{vividviolet}{rgb}{0.62, 0.0, 1.0}
\definecolor{amaranth}{rgb}{0.9, 0.17, 0.31}
\definecolor{palatinateblue}{rgb}{0.15, 0.23, 0.89}
\definecolor{brightpink}{rgb}{1.0, 0.0, 0.5}
\definecolor{cornflowerblue}{rgb}{0.39, 0.58, 0.93}
\definecolor{deepcarminepink}{rgb}{0.94, 0.19, 0.22}
\definecolor{radicalred}{rgb}{1.0, 0.21, 0.37}

\def\H0{{\text{H}\hspace*{-2.05mm}\text{H} 0\hspace*{-1.35mm}0\ }}

\def\be{\begin{equation}}
\def\ee{\end{equation}}
\def\beq{\begin{equation}}
\def\eeq{\end{equation}}
\def\bea{\begin{eqnarray}}
\def\eea{\end{eqnarray}}
\newcommand{\dd}{\textrm{d}}

\begin{document}

\title{High Redshift $\Lambda$CDM Cosmology: To Bin or not to Bin?}

\author{Eoin \'O Colg\'ain}\email{eoin.ocolgain@atu.ie}
\affiliation{Atlantic Technological University, Ash Lane, Sligo, Ireland}
\author{M. M. Sheikh-Jabbari}\email{jabbari@theory.ipm.ac.ir}
\affiliation{School of Physics, Institute for Research in Fundamental Sciences (IPM), P.O.Box 19395-5531, Tehran, Iran}
\author{Rance Solomon}\email{rancesol@buffalo.edu}
\affiliation{HEPCOS, Department of Physics, SUNY at Buffalo, Buffalo, NY 14260-1500, USA}

\begin{abstract}
We construct observational Hubble $H(z)$ and angular diameter distance $D_{A}(z)$ mock data collected in redshift bins with baseline Planck $\Lambda$CDM input values, before fitting the $\Lambda$CDM model to study evolution of probability density functions (PDFs) of best fit cosmological parameters $(H_0, \Omega_m, \Omega_k)$ across the bins. We find that PDF peaks only agree with the input parameters in low redshift ($z \lesssim 1$) bins for $H(z)$ and $D_{A}(z)$ constraints, and in all redshift bins when $H(z)$ and $D_{A}(z)$ constraints are combined. When input parameters are not recovered, we observe that PDFs exhibit non-Gaussian tails towards larger $\Omega_m$ values and shifts to (less pronounced) peaks at smaller $\Omega_m$ values. This flattening of the PDF is expected as $H(z)$ and $D_{A}(z)$ observations only constrain combinations of cosmological parameters at higher redshifts, so uniform PDFs are expected. Our analysis leaves us with a choice to bin high redshift data in the knowledge that it may be unlikely to recover Planck values, or conduct full sample analysis that biases $\Lambda$CDM inferences to the lower redshift Universe.

\end{abstract}

\maketitle

\section{Introduction} 
Discrepancies between early and late Universe inferences of cosmological parameters, notably the Hubble constant $H_0$ \cite{Riess:2021jrx, Planck:2018vyg} and the $S_8 := \sigma_8 \sqrt{\Omega_m/0.3}$  parameter \footnote{$\Omega_m$ is the matter density today ($z=0$) and $\sigma_8$ is the linear theory amplitude of matter fluctuations averaged in spheres of radius $8h^{-1}$ Mpc with $h:= H_0/100$.} \cite{Planck:2018vyg, Heymans:2020gsg, DES:2021wwk}, suggest that standard model of cosmology may be in crisis (see \cite{Verde:2019ivm, DiValentino:2021izs, Perivolaropoulos:2021jda, Abdalla:2022yfr} for reviews). When in any semblance of crisis, it is invariably instructive to return to the basics. Starting from the Friedmann equations, it is a simple analytic inference that one expects evolution of $H_0$ in redshift ranges when any FLRW cosmology breaks down \cite{Krishnan:2020vaf}. Thus, if $H_0$ and $S_8$ tensions  are due to model breakdown and not systematics, one should expect redshift evolution of $H_0$  in \textit{some} redshift range within the flat $\Lambda$CDM model. Moreover, if redshift evolution \textit{happens in the late Universe}, once $H_0$ evolves, so too must  $\Omega_m$, in which case, one expects $S_8$ to be impacted. Hints of this expected (if one buys $H_0$ and $S_8$ tensions) redshift evolution may already exist in the literature \cite{Risaliti:2018reu, Wong:2019kwg, Millon:2019slk, Krishnan:2020obg, Lusso:2020pdb, Dainotti:2021pqg, Dainotti:2022bzg, Schiavone:2022shz}. Finally, note that the ``early versus late Universe" tensions narrative \cite{Verde:2019ivm} is essentially a redshift evolution narrative.  

In this letter we continue earlier investigations \cite{Colgain:2022nlb, Colgain:2022rxy} of redshift evolution of $\Lambda$CDM cosmological parameters in order to establish the extent to which it is tolerated. We assume Gaussian statistics through least squares fitting and work with Gaussian mock data based on Dark Energy Spectroscopic Instrument (DESI) forecasts \cite{DESI:2016fyo} and baseline Planck-$\Lambda$CDM input parameters \cite{Planck:2018vyg}. Throughout, we document the extent to which the input parameters are recovered in distributions of best fit cosmological parameters as the effective redshift of the mock sample varies.  In recent years, the possibility of non-zero  $\Omega_k$ values and the implications have attracted considerable attention \cite{ Handley:2019tkm, DiValentino:2019qzk, Efstathiou:2020wem, DiValentino:2020hov, DES:2022ygi,Semenaite:2022unt, Yang:2022kho, Jesus:2019jvk, Handley:2019anl, DiValentino:2019dzu, Wang:2019yob, Velasquez-Toribio:2020had, Heinesen:2020sre, Noh:2020vnk, Gao:2020irn, Abbassi:2020drc, Bose:2020cjb, Handley:2020hdp, Nunes:2020uex, Liu:2020pfa, DiValentino:2020srs, Chudaykin:2020ghx, Shimon:2020dvb, Vagnozzi:2020rcz, Gordon:2020gel, KiDS:2020ghu, DiValentino:2020kpf, Qi:2020rmm, Vagnozzi:2020dfn, Cespedes:2020xpn, Zheng:2020tau, Yang:2021hxg, Cao:2021ldv, Zhang:2021djh, Benetti:2021div, Arjona:2021hmg, Acquaviva:2021jov, Dhawan:2021mel, Gonzalez:2021ojp,Oztas:2021kka, Ryan:2021eiw, Zhao:2021jeb, Qi:2022sxm, Wang:2022rvf, He:2021rzc, Geng:2021hqc, Li:2021gab, Zuckerman:2021kgm, Bargiacchi:2021hdp, Kiefer:2021iko, Akarsu:2021max, Shumaylov:2021qje, Adhikari:2022moo, Fondi:2022tfp, Zhang:2022lta, Liu:2022hsu, Glanville:2022xes, Hergt:2022fxk, Baumgartner:2022jdz, Bel:2022iuf, Chatzidakis:2022mpf, Wu:2022fmr, Liu:2022mpj, Luciano:2022ffn}. In particular, it has been noted that Planck prefers a closed Universe, $\Omega_k < 0$ \cite{Planck:2018vyg, Handley:2019tkm, DiValentino:2019qzk, Efstathiou:2020wem, DiValentino:2020hov, DES:2022ygi,Semenaite:2022unt, Yang:2022kho},  so we also study how curvature is impacted as redshift ranges change. 

\section{Set Up} 
We will be interested in both flat and non-flat $\Lambda$CDM models for which  $H(z)$ takes the form, 
\be
\label{hubble_lcdm}
H(z) := H_0 E(z)= H_0 \sqrt{\Omega_{\Lambda} + \Omega_{k} (1+z)^2 + \Omega_{m} (1+z)^3},  
\ee
where $\Omega_{\Lambda} = 1 - \Omega_m - \Omega_k$, so that we recover the Hubble constant $H_0 = H(z=0)$ at $z=0$. We work with low redshift data, so (\ref{hubble_lcdm}) is a justified approximation. Thus, in addition to $H_0$, the model depends on two constants, $\Omega_k$ and $\Omega_{m}$, and we specialise to flat $\Lambda$CDM through $\Omega_k = 0$. In addition, the angular diameter distance $D_{A}(z)$ depends on $\Omega_{k}$: 
\be
\label{diameter_distance}
D_{A}(z) = 
\begin{cases}
       \frac{c}{(1+z) H_0 \sqrt{\Omega_k}} \sinh \left( \sqrt{\Omega_k} {\cal X}(z)  \right)  & \Omega_k > 0 \\
      \frac{c {\cal X}(z) }{(1+z)  H_0}  & \Omega_k = 0 \\
       \frac{c}{(1+z) H_0 \sqrt{-\Omega_k}} \sin \left( \sqrt{-\Omega_k} {\cal X}(z) \right)  & \Omega_k <  0,
    \end{cases}  
\ee
where ${\cal X}(z):= \int_0^{z} \frac{\dd z^{\prime}}{E(z^{\prime})}$. 
The mock data we use is based on DESI forecasts assuming $14,000$ deg$^2$ coverage \cite{DESI:2016fyo}. The forecast comprises 29 $H(z)$ and 29 $D_{A}(z)$ determinations in the redshift range $0.05 \leq z \leq 3.55$, which following \cite{Colgain:2022rxy}, we split into four redshift bins with similar data quality in each bin: $0.05 \leq z < 0.8$, $0.8 < z < 1.5$, $1.5 < z < 2.3$ and $2.3 < z \leq 3.55$. Throughout we construct mock $H(z)$ and $D_{A}(z)$ constraints based on the baseline Planck-$\Lambda$CDM model ($\Omega_k = 0$), typically 10,000 realisations at a time, before fitting back the model to each realisation using least squares fitting. More concretely, our mock data points are generated in normal distributions about the baseline cosmology using the forecasted DESI errors \cite{DESI:2016fyo}. This exercise allows us to construct a distribution of best fit parameters in each redshift bin and check if the input parameters are recovered.    

\section{Flat $\Lambda$CDM}
As a warm-up, we revisit the analysis presented in \cite{Colgain:2022rxy}. Therein, we worked with mock DESI data, but only $H(z)$ constraints, and exclusively the flat $\Lambda$CDM model, where we imposed an appropriate Gaussian Planck prior, $\Omega_{m} h^2 = 0.1430 \pm 0.0011$ \cite{Planck:2018vyg} with $h:=H_0/100$. Observe that since (\ref{hubble_lcdm}) scales as $H(z) \propto \sqrt{ \Omega_m h^2 (1+z)^3}$ for large $z$, this prior guides the high redshift behaviour of $H(z)$. Here we make two further generalisations. First, we include forecasted DESI $D_{A}(z)$ constraints and secondly relax the high redshift Gaussian prior on $\Omega_{m} h^2$ so that it is more agnostic about the late Universe, $\Omega_m h^2 = 0.141 \pm 0.006$ \cite{Vonlanthen:2010cd, Verde:2016wmz} (see appendix in \cite{Krishnan:2021dyb}). The precise form of this prior is not so important, simply that it exists and does not restrict us to the baseline Planck-$\Lambda$CDM model from the outset. As should be clear from a comparison of the errors, our new $\Omega_{m} h^2$ prior has been relaxed, but it still offers guidance for the high redshift behaviour of $H(z)$. We display results both with and without this prior. 

\begin{figure}[htb]
\includegraphics[width=80mm]{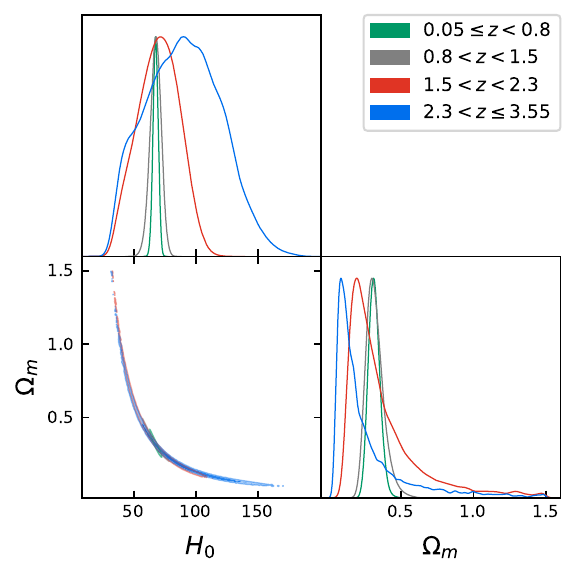} 
\caption{Distributions of approximately 10,000 best fits of the flat $\Lambda$CDM model to Planck-$\Lambda$CDM mocks of $H(z)$ data. No prior on $\Omega_m h^2$ is imposed. Best fits that fall outside of the range $0 < \Omega_m  < 1.5$ have been removed. The key features are non-Gaussian tails towards larger $\Omega_m$ values and a shift in the peak of the best fit distribution to lower $\Omega_m$ values as the effective redshift of the mock data increases.}
\label{fig:H_no_prior}
\end{figure}

Our methodology follows \cite{Colgain:2022rxy}. We focus on the four redshift ranges outlined earlier and in each range or bin construct mock DESI $H(z)$ and $D_{A}(z)$ data based on the baseline Planck-$\Lambda$CDM cosmology, $H_0 = 67.36$ km/s/Mpc, $\Omega_{m} = 0.3153$ and $\Omega_k = 0$. Thus, we assume that the Universe is flat, but as we shall see in section \ref{non-flat_lcdm}, redshift evolution can lead to different inferences. In each bin, we mock up data and then fit back either the flat $\Lambda$CDM model, or in the next section, the non-flat $\Lambda$CDM model. We begin by focusing on $H(z)$ mocks, which allow us to recover the results in \cite{Colgain:2022rxy}. 

\begin{figure}[htb]
\includegraphics[width=80mm]{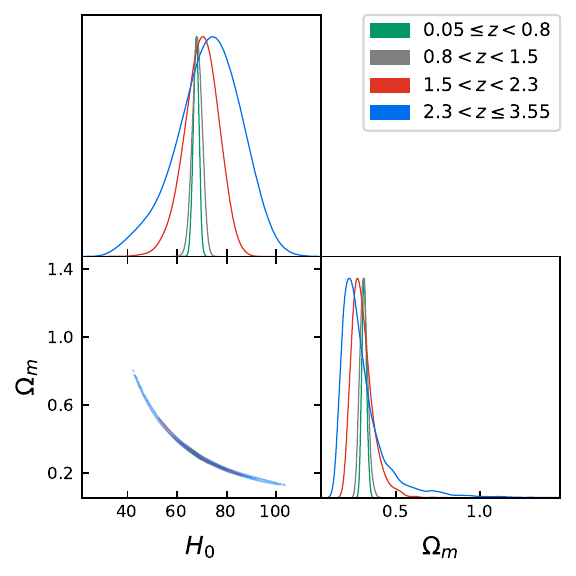} 
\caption{Same as Fig. \ref{fig:H_no_prior}, but with the introduction of a Gaussian prior $\Omega_m h^2 = 0.141 \pm 0.006$, which simply makes the distributions narrower.}
\label{fig:H_prior}
\end{figure}

In Fig. \ref{fig:H_no_prior} we show the result of fitting back the flat $\Lambda$CDM model to a large number of mocks of $H(z)$ DESI data as the redshift range is varied. For visualisation purposes, we have not normalised the PDFs, but comment on this below. In practice, we allow generous uniform priors, $H_0 \in [0, \infty)$ km/s/Mpc and $\Omega_{m} \in [0, 1.5]$, but for visualisation purposes we restrict our attention to best fits in the range $0 < \Omega_{m} < 1.5$,\footnote{The choice of upper bound here is arbitrary and can be changed without affecting conclusions. In particular, the non-Gaussian tails are still there for arbitrarily large upper bounds on $\Omega_m$.} thereby removing any best fits that saturate (or violate) the bounds.  This means that in contrast to \cite{Colgain:2022rxy}, the distributions we observe are not impacted by the bounds. For a number of mocks, as an added consistency check, we confirmed that extremisation of likelihood to get a best fit agrees with the median from an Markov Chain Monte Carlo chain, so there is nothing to indicate that we are not correctly identifying best fit parameters. The main take-aways are that as one increases the redshift range, in line with expectations, the errors on $H_0$ and $\Omega_{m}$ increase, however $\Omega_{m}$ best fits develop non-Gaussian tails in the direction of larger $\Omega_{m}$ values, while $H_0$ best fits develop non-Gaussian tails towards smaller $H_0$ values. (This non-Gaussian tail is more visible in Fig.\ref{fig:H_prior}.) Here, recall that $H_0$ and $\Omega_m$ are anti-correlated.\footnote{One notable exception to this is galaxy BAO, where variations in the radius of the sound horizon at drag epoch lead to $H_0$ and $\Omega_m$ being positively correlated at lower redshifts \cite{Addison:2017fdm}.} Finally, it is worth noting that the peaks of the distributions move to smaller $\Omega_{m}$ and larger $H_0$ values as the redshift range is increased. As shown in \cite{Colgain:2022rxy}, the peaks also become less pronounced and this is expected because the PDF should produce a uniform PDF at high redshift. Throughout, the expected input cosmological parameters are recovered in the lower redshift bins. 

While the flattening of the PDF is expected, because $H(z)$ should only be sensitive to the combination $\Omega_m h^2$ at very high redshifts, thereby leaving $H_0$ and $\Omega_m$ unconstrained, the non-Gaussian tails require some explanation. Let us define $A:=H_0^2 (1-\Omega_m)$ and $B:=H_0^2 \Omega_m$. As explained, $B$ is well constrained by $H(z)$ data at high redshifts, however $A$ is not, so any distribution in $A$ must spread, namely the uncertainty in $A$ must increase. Noting that $\Omega_m \geq 0$, $A$ can only spread effectively to larger values through increasing $H_0$. If $H_0$ increases, then $\Omega_m$ decreases because $B$ is constrained, and this explains the shift in the peaks. However, we still need $A$ to spread to smaller values and this is achieved through non-Gaussian tails in the directions of larger $\Omega_m$ and smaller $H_0$ values. Throughout this process, the central value of the $A$ distribution changes little, so the shifted peaks and non-Gaussian tails maintain the original mean, as can be seen by inspection in Fig. \ref{fig:H_no_prior} and Fig. \ref{fig:H_prior}. Observe that the non-Gaussian tails are more developed with the prior than without, especially for $H_0$. 

Next, we repeat the exercise, but introduce a Gaussian prior on $\Omega_m h^2$ in order to document any effect. The result of this exercise is shown in Fig. \ref{fig:H_prior}. As may be expected, one finds narrower distributions, but the plot is qualitatively the same as Fig. \ref{fig:H_no_prior}. It is worth stressing that previous analysis \cite{Colgain:2022rxy} focused on $H(z)$ DESI mocks subject to a more stringent Gaussian prior on $\Omega_{m} h^2$. Here we have relaxed the prior, but the features are robust; we once again see that the peak of the distribution of best fits moves in the direction of $\Omega_m = 0$, while non-Gaussian tails in the direction of larger $\Omega_m$ values appear.  The fact that the introduction of the $\Omega_m h^2$ prior does not greatly impact the plot is understandable when one recalls that high redshift $H(z)$ constraints are largely only sensitive to the combination $\Omega_m h^2$, so one has similar information in the $H(z)$ constraints and $\Omega_m h^2$ prior, the effect of which is simply to make the distribution of best fits narrower. 

\begin{figure}[htb]
\includegraphics[width=80mm]{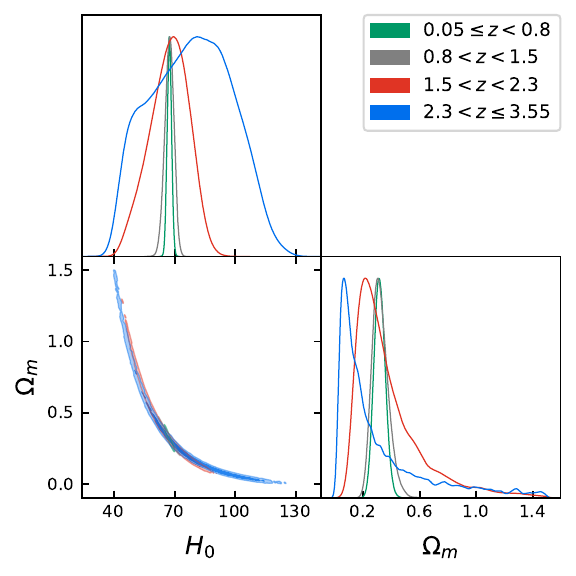} 
\caption{Same as Fig. \ref{fig:H_no_prior}, but $H(z)$ constraints replaced by $D_{A}(z)$ constraints.}
\label{fig:DA_no_prior}
\end{figure}

\begin{figure}[htb]
\includegraphics[width=80mm]{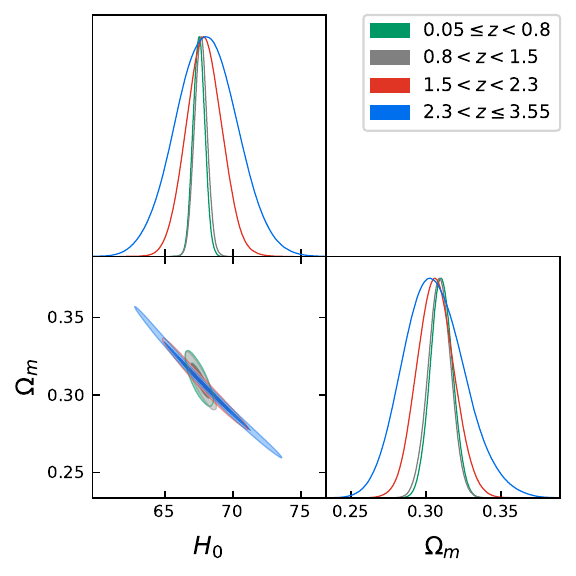} 
\caption{Same as Fig. \ref{fig:H_prior}, but $H(z)$ constraints replaced by $D_{A}(z)$ constraints. In contrast to Fig. \ref{fig:H_no_prior}, \ref{fig:H_prior} and \ref{fig:DA_no_prior}, distributions of best fit parameters are approximately Gaussian in all redshift ranges.}
\label{fig:DA_prior}
\end{figure}

\begin{figure}[htb]
\includegraphics[width=80mm]{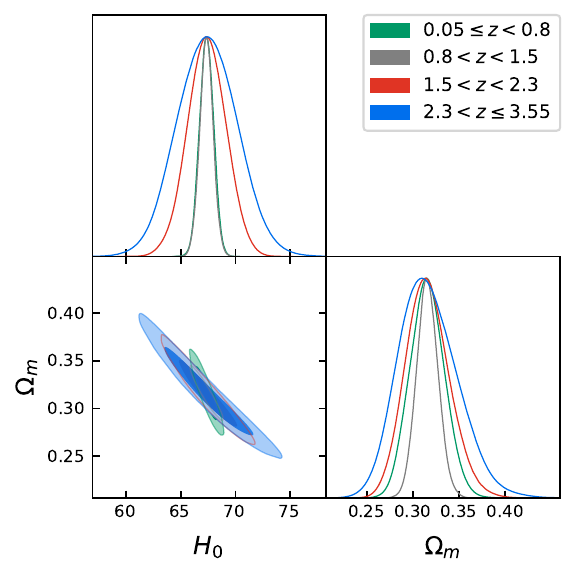} 
\caption{Distributions of approximately 10,000 best fits of the flat $\Lambda$CDM model to Planck-$\Lambda$CDM mocks of $H(z)$ and $D_{A}(z)$ data. In the same fashion as Fig. \ref{fig:DA_prior}, the distributions are Gaussian in all redshift ranges and any evolution in the peak of the distribution is negligible.}
\label{fig:H_DA_no_prior}
\end{figure}

Let us now switch attention to $D_{A}(z)$ mocks, where we will see that the introduction of a $\Omega_m h^2$ prior radically changes the conclusion. The result of fitting $D_{A}(z)$ constraints without an $\Omega_m h^2$ prior is shown in Fig. \ref{fig:DA_no_prior}, which is more or less the same as Fig. \ref{fig:H_no_prior} and Fig. \ref{fig:H_prior}. Once again, a shift in the peak of the distribution and non-Gaussian tails are evident at higher redshifts. The peak shifts and non-Gaussian tails can once again be explained, but in terms of the $A= H_0^2 (1-\Omega_m)$ distribution being constrained, while $B = H_0^2 \Omega_m$ is forced to spread. It should be noted that one sees the same effect for mock Type Ia supernovae (SN) data \cite{Colgain:2022nlb} and this result is expected since luminosity distance $D_{L}(z)$ and angular diameter distance $D_{A}(z)$ are related, $D_{L}(z) = (1+z)^2 D_{A}(z)$. Nevertheless, the introduction of a prior makes a pronounced difference, as is evident from Fig. \ref{fig:DA_prior}. A slight shift in the peak of the distribution is evident, but the non-Gaussian tails have disappeared in all redshift ranges. Note, since $H(z)$ observations constrain the combination $\Omega_m h^2$ well at higher redshifts, one expects that combining $D_{A}(z)$ and $H(z)$ constraints will lead to regular Gaussians. This turns out to be the case as is evident from Fig. \ref{fig:H_DA_no_prior}. 

This result just needs interpretation. There is an obvious difference between $H(z)$ and $D_{A}(z)$ constraints in the sense that the former is a direct constraint on the Hubble parameter, whereas the latter constrains an integral of the inverse of the Hubble parameter. Indeed, as is clear from the $\Omega_k = 0$ entry in (\ref{diameter_distance}), since $H(z)$ is a strictly increasing function with redshift, this integral is more sensitive to variations in $H(z)$ near $z=0$ and becomes less sensitive as $z$ increases. This is a simple mathematical fact. In essence, what the evolution in Fig. \ref{fig:DA_no_prior} is telling us is that neither $H_0$ nor $\Omega_m$ are well determined at higher redshifts when one uses $D_{A}(z)$ constraints on their own; the nature of the integral implies that $D_{A}(z)$ is largely only sensitive to the combination $\Omega_{\Lambda} h^2 = (1-\Omega_m) h^2$ in high redshift bins. This result immediately extends to $D_{L}(z)$ constraints and SN cosmology. However, the introduction of the prior $\Omega_m h^2$, which is relevant at higher redshifts, allows one to fully recover the input parameters and prevent any evolution. It is then a simple deduction that we should get the same effect by combining $D_{A}(z)$ and $H(z)$ constraints, since the latter only contain information about $\Omega_m h^2$ in higher redshift bins. In appendix \ref{sec:HvsDA}, we make a like for like comparison between $H(z)$ and $D_{A}(z)$ constraints showing the differences in sensitivity at low and high redshift.


\section{Non-flat $\Lambda$CDM}
\label{non-flat_lcdm}
\begin{figure}[htb]
\includegraphics[width=80mm]{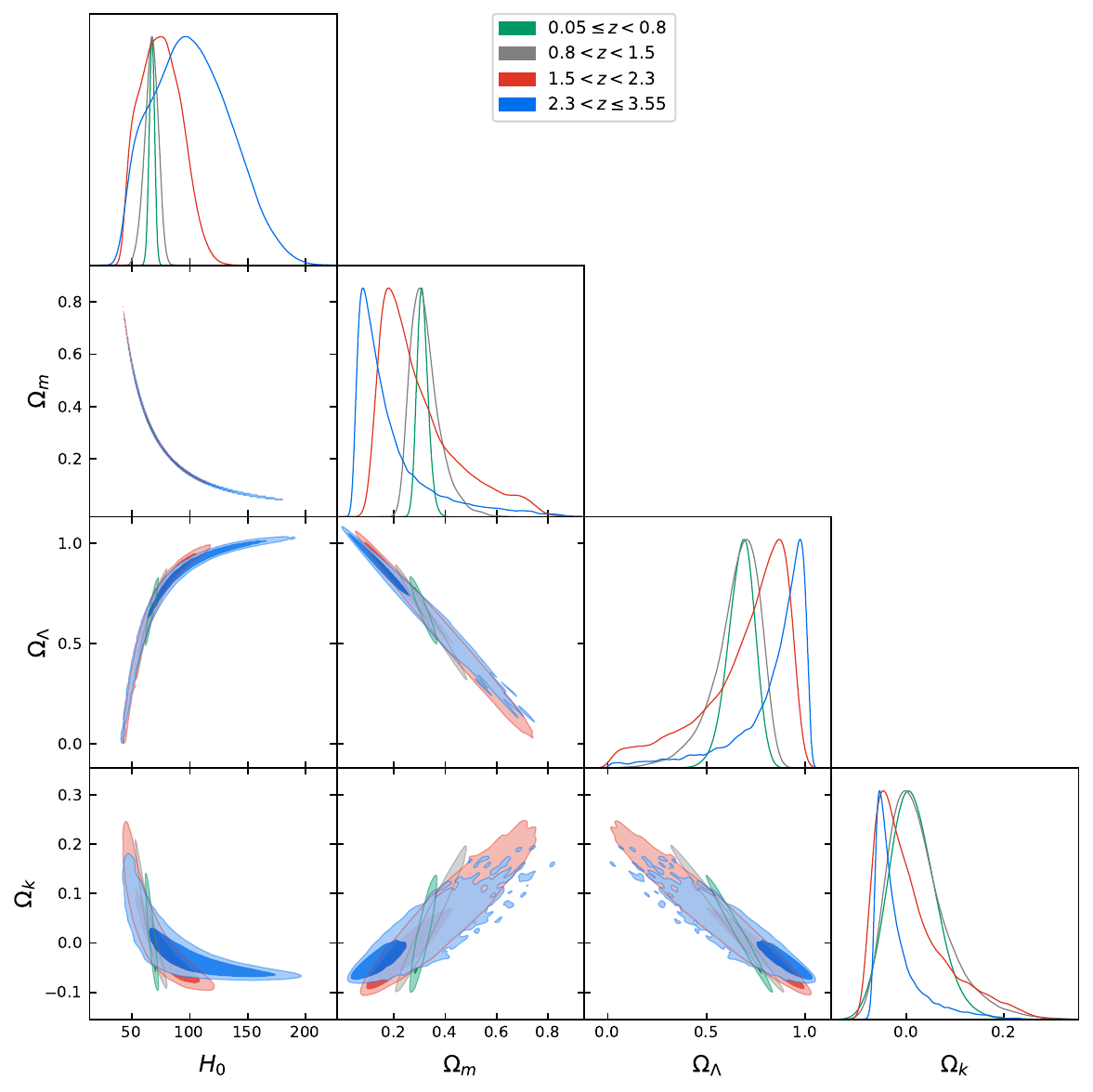} 
\caption{Same as Fig. \ref{fig:H_prior} but extended from the flat $\Lambda$CDM model to the non-flat $\Lambda$CDM model. The combination of $H(z)$ constraints with a $\Omega_m h^2 = 0.141 \pm 0.006$ prior does not preclude evolution and non-Gaussian tails. Negative values of curvature, $\Omega_k < 0$, are preferred at higher redshifts as is evident from the shift in the peak of best fit $\Omega_k$ values to negative values. }
\label{fig:curvature_H_prior}
\end{figure}

Introducing curvature in our analysis is straightforward. The only issue is that we have an extra parameter, so this means that the mock data is less constraining with the three parameters $(H_0, \Omega_{m}, \Omega_{k})$. For this reason, we will impose the Planck prior on $\Omega_m h^2$ in all the fits from the outset. What this does is essentially fix the high redshift behaviour, so that evolution only happens between the dark energy and curvature sectors. This can be seen by converting the array of $(H_0, \Omega_{m}, \Omega_{k})$ best fits into $A := H_0^2 \Omega_{\Lambda} = H_0^2 (1-\Omega_m - \Omega_k)$, $B := H_0^2 \Omega_m$ and $C := H_0^2 \Omega_k$ and noting that $B$ distributions remain Gaussian throughout, where they are constrained by the prior. Incidentally, the $\Omega_m h^2$ prior also ensures that both $A$ and $B$ are Gaussian when $C = \Omega_k = 0$, but without the prior, the distributions $(A, B, C)$ are generically not Gaussian. We leave this exercise to the reader. In Fig. \ref{fig:curvature_H_prior} we show the distributions of best fit non-flat $\Lambda$CDM model parameters, where in each case we mock up on the baseline Planck-$\Lambda$CDM cosmology with $\Omega_k = 0$. Note, here we have imposed uniform priors $H_0 \in [0, \infty), \Omega_m \in [0, 1.5], \Omega_{\Lambda} \in [0,1.5]$, but have removed all configurations that saturate these bounds in the final configurations, thereby leaving Planck-$\Lambda$CDM mocks with best-fits within these priors. We have checked that adjusting the priors to allow $\Omega_\Lambda < 0$ does not change results. Once again, the upper bounds represent an arbitrary choice, which can be changed. 

\begin{figure}[htb]
\includegraphics[width=80mm]{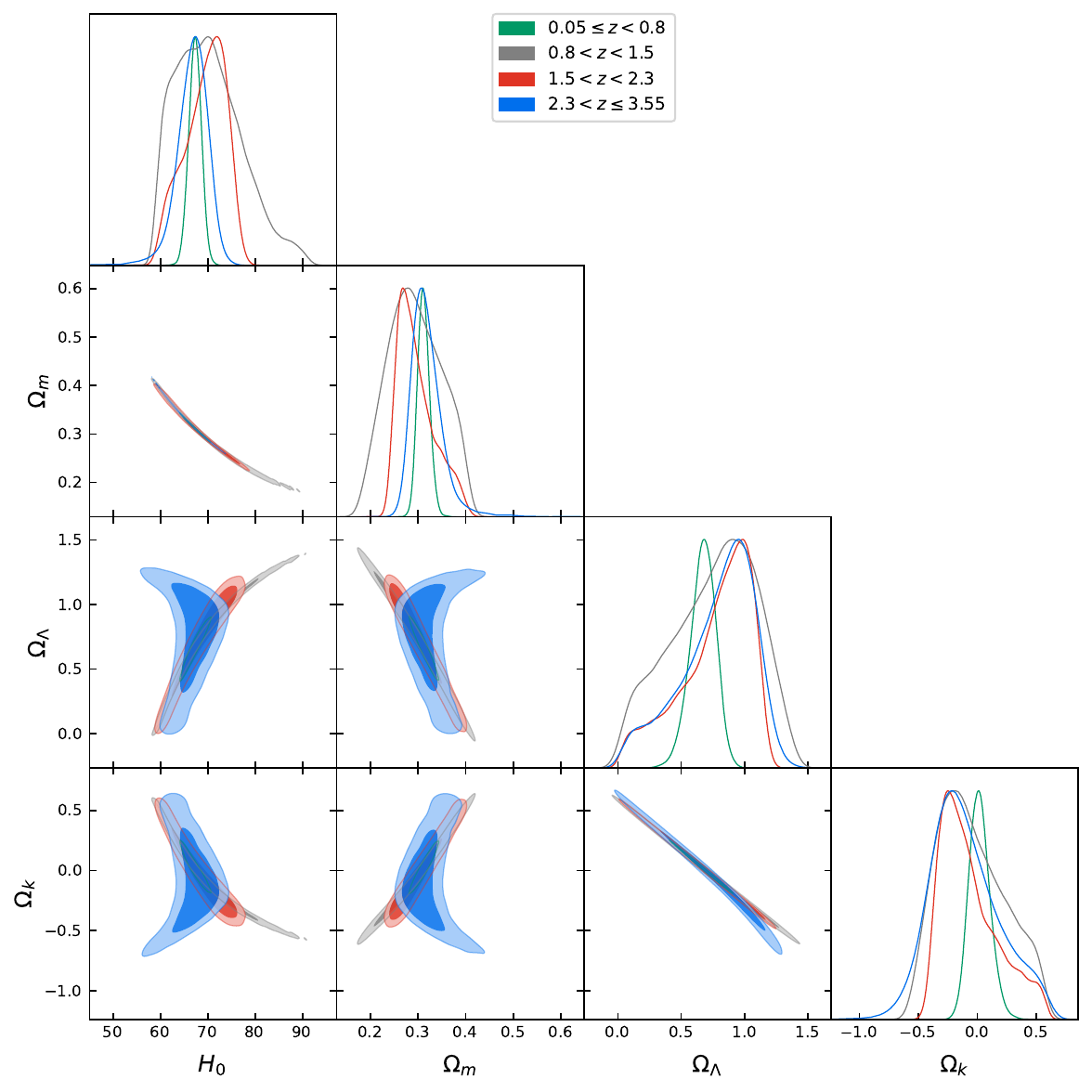} 
\caption{Same as Fig. \ref{fig:DA_prior}, but $H(z)$ constraints replaced with $D_{A}(z)$ constraints.}
\label{fig:curvature_DA_prior}
\end{figure}

\begin{figure}[htb]
\includegraphics[width=80mm]{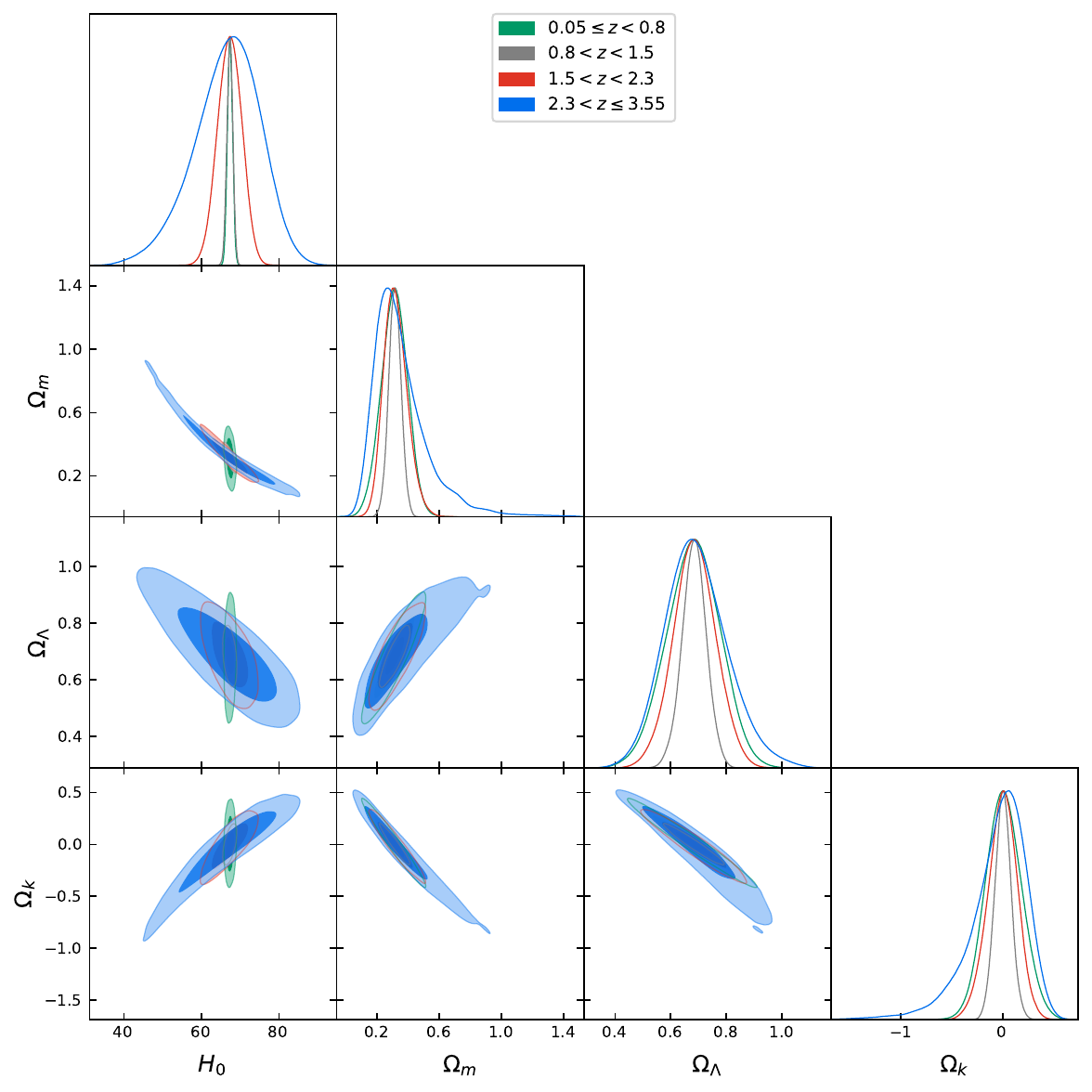} 
\caption{Same as Fig. \ref{fig:H_DA_no_prior} but with curvature parameter $\Omega_k$.}
\label{fig:curvature_H_DA_no_prior}
\end{figure}

In line with expectations from the last section, we see the emergence of non-Gaussian tails in the direction of larger $\Omega_m$, smaller $\Omega_{\Lambda}$ values and positive $\Omega_k$ values in higher redshift bins for both $H(z)$ and $D_{A}(z)$ mocks in Fig. \ref{fig:curvature_H_prior} and Fig. \ref{fig:curvature_DA_prior}, respectively. In line with earlier results from our flat $\Lambda$CDM analysis, this effect is less pronounced when one combines $D_{A}(z)$ constraints with a prior on $\Omega_{m} h^2$. Nevertheless, it is worth noting that the peak of the distributions moves towards smaller $\Omega_m$ values, larger $\Omega_{\Lambda}$ values, and most interestingly, negative $\Omega_k$ values (recall the sum rule $\Omega_k+\Omega_\Lambda+\Omega_m=1$), despite all mocks being consistent with baseline Planck-$\Lambda$CDM with $\Omega_k = 0$. In Fig. \ref{fig:curvature_H_DA_no_prior} we drop the $\Omega_m h^2$ prior and combine $H(z)$ and $D_{A}(z)$ constraints. Interestingly, this succeeds in recovering the input parameters and very little evolution is evident. In particular, one finds that $\Omega_k=0$. 

\section{Discussion}
Our mocking, binning and fitting procedure for flat $\Lambda$CDM has led to both expected and unexpected results. In particular, we recover mock input cosmological parameters from $H(z)$ and $D_{A}(z)$ constraints in low redshift bins, as well as from combinations of $H(z)$ and $D_{A}(z)$ constraints in all redshift bins. However, we record shifts in the peaks of best fit distributions and non-Gaussian tails for $H(z)$ and $D_{A}(z)$ constraints as the effective redshift increases. From our distribution or PDF we can extract a probability or $p$-value for finding a best fit in any given $H_0$, $\Omega_m$ or $\Omega_k$ interval. The location of the peak and existence of non-Gaussian tails have a bearing on these $p$-values. We have removed all configurations that saturate or violate the bounds, so our priors have no bearing on results, \textit{cf.} \cite{Colgain:2022rxy}. Here, we noted that $H(z)$ is more sensitive to the combination $\Omega_m h^2$ at higher redshifts, whereas $D_{A}(z)$, being an integral of the inverse of $H(z)$, is more sensitive to $\Omega_{\Lambda} h^2 = (1-\Omega_m) h^2$ at higher effective redshifts. In short, both $H(z)$ and $D_{A}(z)$ can evidently distinguish cosmological parameters in low redshift bins, but can no longer distinguish parameters as well in high redshift bins, unless combined, whereby the constraints complement each other, e. g. anisotropic BAO \cite{BOSS:2016wmc, Bautista:2020ahg, Gil-Marin:2020bct, Hou:2020rse, Neveux:2020voa, duMasdesBourboux:2020pck}.

It is good to dwell a bit on these results to elucidate them further. Recall that $H(z)$ at very high redshift ($z \gg 1$), yet in the late Universe (we ignore radiation), only constrains the combination $\Omega_m h^2$. This leaves $H_0$ and $\Omega_m$ unconstrained, so in very high redshift bins, one expects uniform PDFs whereby any value of $H_0$ or $\Omega_m$ is equally probable. Similar logic applies to $D_{A}(z)$ constraints. Our mock Planck-$\Lambda$CDM analysis shows that Gaussian PDFs at low redshift develop non-Gaussian tails towards larger $\Omega_m$ values and shifts in peaks to smaller $\Omega_m$ values at intermediary redshifts (see Fig. \ref{fig:H_no_prior} and Fig. \ref{fig:H_prior} for $H(z)$ constraints). As highlighted in \cite{Colgain:2022rxy} (see appendix), when correctly normalised, the peak of the PDF decreases in prominence as it moves to smaller $\Omega_m$ values, so evidently the PDF is approaching the expected uniform PDF. Nevertheless, throughout this process, we retain cosmological information about our mock data. Surprisingly, \textit{one arrives at the conclusion that even if one prepares a large number of Planck-$\Lambda$CDM universes with $\Omega_m \sim 0.3$, one does not expect to recover $\Omega_m \sim 0.3$ in high redshift bins with exclusively $H(z)$ or $D_{A}(z)$ data.} This applies to both cosmic chronometer \cite{Jimenez:2001gg, Moresco:2016mzx} and SN samples \cite{SDSS:2014iwm, Pan-STARRS1:2017jku, Brout:2022vxf}, since $D_{L}(z) \propto D_{A}(z)$. Interestingly, existing results point to redshift evolution of cosmological parameters in observed (or real) $H(z)$ and $D_{L}(z)$ data with increasing effective redshift \cite{Risaliti:2018reu, Krishnan:2020obg, Lusso:2020pdb, Dainotti:2021pqg, Dainotti:2022bzg, Schiavone:2022shz}.

Thus, different values of cosmological parameters are expected in high redshift data sets. Tellingly, standardisable QSOs already return unexpectedly large $\Omega_m$ values \cite{Risaliti:2018reu, Lusso:2020pdb} and recent JWST $\Lambda$CDM anomalies \cite{Boylan-Kolchin:2022kae, Menci:2022wia, Haslbauer:2022vnq} also hint at departures from Planck-$\Lambda$CDM expectations at high redshifts. Unfortunately, how we currently analyse late Universe samples is heavily biased towards the low redshifts, $z \lesssim 1$, where parameters can be more easily uniquely determined. What our analysis shows is that there is cosmological information in high redshift $H(z)$ and $D_{A}(z)$ constraints, which can be unlocked. As explained, one reaches an extreme high redshift where only combinations of parameters are constrained, and this is described by uniform PDFs, but none of our PDFs are uniform and are simply approaching such an outcome, so they clearly retain information. 

It is standard practice to combine high redshift data, which can less cleanly disentangle cosmological parameters, with low redshift data, where the parameter constraints are much stronger. This reduces the role of high redshift data to essentially a spectator whereby it struggles to influence central values, but only improves errors. In essence, we risk inadvertently blinding ourselves to the high redshift Universe, and potential departures from Planck-$\Lambda$CDM, and the only way to avoid this conclusion is to bin $H(z)$ and $D_{A}/D_{L}(z)$ constraints by redshift range and not simply fit large, expansive samples, as is common practice. Nevertheless, if one bins, one should not expect Planck values from the outset. 

Finally, our analysis shows that redshift evolution of best fit cosmological parameters persists when one includes curvature $\Omega_k$ and confronts the $\Lambda$CDM model with either $H(z)$ or $D_{A}(z)$ constraints either with or without a Gaussian prior on $\Omega_m h^2$. Interestingly, we find a preference for $\Omega_k < 0$ at higher redshifts, despite mocking up on the Planck-$\Lambda$CDM model with $\Omega_k = 0$. Nevertheless, when we combine $D_{A}(z)$ and $H(z)$ constraints, the peak of the distributions agree well with the Planck-$\Lambda$CDM input parameters across all redshifts. This is extremely curious, because there is a puzzling tension in the curvature parameter between (anisotropic) BAO and CMB \cite{Planck:2018vyg, Handley:2019tkm, DiValentino:2019qzk, Efstathiou:2020wem, DiValentino:2020hov, DES:2022ygi, Semenaite:2022unt, Yang:2022kho}. Since curvature is difficult to square with inflation, the debate has spilled over recently into the realms of philosophy and theory \cite{Anselmi:2022uvj, Jimenez:2022asc}, but redshift evolution may yet help us reconcile these conflicting observations provided we take redshift binning seriously.

\section{Acknowledgements}
We thank Stephen Appleby, Eleonora Di Valentino, Avi Loeb and Sunny Vagnozzi for correspondence on related topics. E\'OC thanks the organisers of ``Challenges to $\Lambda$CDM", September 4-7 2022, 
Aristotle University of Thessaloniki,  for a stimulating environment during early stages of this project. MMShJ is supported in part by SarAmadan grant No ISEF/M/401332, and acknowledges support from the ICTP through the Senior Associates Programme (2022-2027).

\appendix

\section{$H(z)$ versus $D_{A}(z)$ constraining power}
\label{sec:HvsDA}

\begin{figure}[h]
\includegraphics[width=80mm]{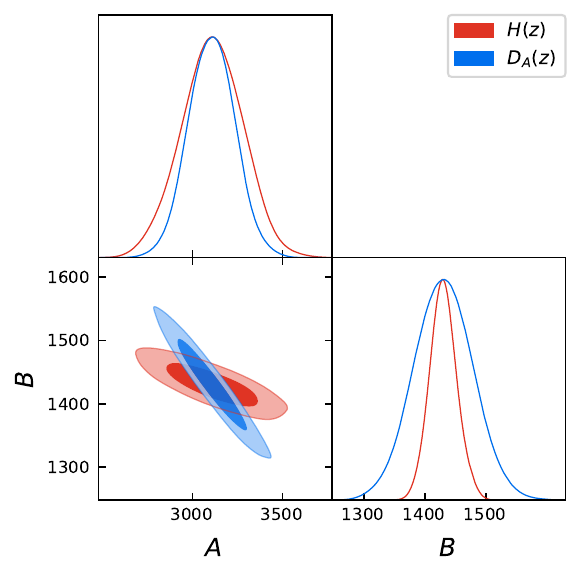} 
\caption{Differences in constraining power in $H(z)$ and $D_{A}(z)$ constraints in low and high redshift regimes, where $A = H_0^2 (1-\Omega_m)$ and $B = H_0^2 \Omega_m$ are the relevant parameters. $H(z)$ constrains $B$ better than $D_{A}(z)$, while $D_{A}(z)$ constrains $A$ better than $H(z)$.}
\label{fig:HvsDA}
\end{figure}

Here we utilise the DESI forecasts over the full range, $0.05 \leq z \leq 3.55$, generate a large number of Planck-$\Lambda$CDM mocks before fitting back the flat $\Lambda$CDM model. This gives us an array of $(H_0, \Omega_m)$ values, which we convert to an array of $A = H_0^2 (1-\Omega_m)$ and $B = H_0^2 \Omega_m$ values. Here $A$ is relevant at low redshifts, whereas B is relevant at high redshifts. The constraints on $A$ and $B$ can be seen in Fig. \ref{fig:HvsDA}, where it is evident that $H(z)$ provides stronger constraints on $B$, while $D_{A}(z)$ provides stronger constraints on $A$. This underscores the complementarity in sensitivity in different redshift regimes.

\end{document}